\documentclass[12pt]{iopart}
\usepackage[utf8]{inputenc}
\usepackage[T2A]{fontenc}
\usepackage{hyphenat}
\usepackage{indentfirst}
\usepackage{graphicx}
\usepackage{float}
\usepackage{dcolumn}
\usepackage{bm}
\usepackage{color}
\usepackage{hyperref}
\usepackage{xcolor}
\usepackage{epsfig}
\usepackage{epstopdf}
\usepackage[margin=2cm]{geometry}
\usepackage[numbers,square,sort&compress]{natbib}
\usepackage{multirow}
\usepackage{booktabs}
\usepackage{adjustbox}

\newcommand{\ddtp}[1]{\frac{\partial{#1}}{\partial{t}}}
\newcommand{\dif}{\mathrm{d}}

\newcommand{\const}{\mathrm{const}}

\begin{document}
\paper[LID-QMS analysis with elliptic Gaussian laser beam]{Gas release from metals under irradiation with elliptic Gaussian laser beam during LID-QMS analysis}
\author{Alexander A. Stepanenko$^*$ (Александр А. Степаненко), Yury M. Gasparyan (Юрий М. Гаспарян)}
\address{Plasma Physics Department, National Research Nuclear University MEPhI (Moscow Engineering Physics Institute), Kashirskoe highway, 31, Moscow 115409, Russia}
\address{$^*$ Author to whom correspondence should be addressed.}
\ead{aastepanenko@mephi.ru}

\begin{abstract}
	Laser-induced-desorption quadrupole-mass-spectrometry (LID-QMS) diagnostics is considered as one of the candidate methods for the remote control of tritium inventory in the ITER first wall. Studies involving LID-QMS generally assume the circular shape of the laser spot on the analyzed surface. At the same time, the diagnostics laser source cannot be always positioned so as to irradiate tokamak tiles under normal angles, which results in the laser spot shape differing from the circular one. In this contribution, we analyze the tritium removal under sample irradiation by an elliptic Gaussian laser beam, extending the results of our previous analysis [A.A.~Stepanenko, Yu.M.~Gasparyan, Physica Scripta 99 (8), 085604 (2025)]. The thermal desorption model governing the heat transport and tritium removal from the solid is formulated. The new analytical expression describing the sample temperature dynamics is derived. The developed model is used to examine the impact of the laser beam/spot ellipticity on the tritium desorption from a tungsten sample. The conditions, under which the elliptic beam can be approximated with the circular one for the rapid assessment of the amount of desorbed tritium, are assessed.
\end{abstract}


\section{Introduction\label{sec:intro}}
A deuterium-tritium mixture will be a fuel for future fusion reactors. Accumulation of radioactive tritium in fusion devices is limited due to safety reasons and it should be monitored remotely. One of key mechanisms of fuel accumulation in fusion devices is co-deposition, that is trapping of impinging particles in continuously growing films in deposition dominated areas. To remotely probe the local tritium content in these layers or other subsurface layers, a number of laser-based methods were developed, involving the light- or mass-spectrometry of the desorption or ablation fluxes \cite{oelmann2021analyses,lyu2022characterization,wu2023effect,almaviva2023libs,marenkov2023dynamics,yehia2023considerations,vujadinovic2025hydrogen,zlobinski2024first}. Among these approaches, laser-induced-desorption quadrupole-mass-spectrometry, or LID-QMS, is considered as one of the main candidate methods for the nondestructive control of tritium content in ITER \cite{zlobinski2011laser,de2017efficiency,huber2026innovative}.

The LID-QMS diagnostic capabilities were intensively developed and analyzed over the years \cite{huber2001situ,schweer2007situ,schweer2009laser,zlobinski2011laser,zlobinski2013hydrogen,zoomers2013applicability,van2013laser,de2018temperature, zlobinski2020efficiency,widdowson2021evaluation,lyu2021characterization,wu2023effect,yehia2023considerations,lyu2024study,medvedev2024lia,zlobinski2024first,widdowson2025overview,wang2025efficiency,lyu2025quantitative,matveev2025analysis,he2025towards}. The interpretation of LID-QMS data relies on using numerical simulations of the desorption, including both the particle and heat transport in the sample. The thermal desorption models for LID-QMS frequently assume the one-dimensional approximation for the heat transfer \cite{zlobinski2011laser,yu2017deuterium,matveev2020crds,gasparyan2021laser} and the circular shape of the laser spot found in experiments \cite{zlobinski2011laser,zlobinski2019laser,zlobinski2024first,medvedev2024lia}. 

Recently, it was demonstrated \cite{stepanenko2024dimensional} that the one-dimensional limit for the heat transfer problem is valid only for laser beams having sufficiently large spot area, with the characteristic size being much larger than the heat diffusion length defined at the end of the laser pulse. For laser spots having dimensions comparable to or smaller than the heat diffusion length, two-dimensional effects related to the spatial heat re-distribution must be taken into account leading to a more complex description of the temperature dynamics in the sample. The circular shape of the laser spot can be also distorted, since the laser light sources cannot be always positioned so as to ensure the normal light incidence on the tile surface \cite{xiao2013application,jia2021effect,wu2022spatiotemporal,wu2023effect}. Elliptically elongated spot shapes can be characterized by transverse dimensions that are smaller than the heat diffusion length, also requiring the two-dimensional approach to resolving the heat transport.

Temperature dynamics in solids driven by elliptical heat sources were addressed in a number of studies devoted to laser micro-machining and welding of materials. Nissim et al.~\cite{nissim1980temperature} performed theoretical analysis of the temperature dynamics in semiconductors driven by a scanning elliptical laser beam. The resulting relation for the temperature profile was given as an integral relation. Moody and Hendel~\cite{moody1982temperature} addressed a similar analytical problem, incorporating the non-linear dependencies of the material properties on the sample temperature, and derived relations for the sample temperature also in the form of integral relations. Miyazaki and Giedt \cite{toshiyuki1982heat} worked out the analytical solution to the problem of heat transfer in an infinite plate driven by the elliptical cylindrical heat source. Solutions were derived as infinite series of the Mathieu functions. Goldak \cite{goldak1984new} developed a numerical model of thermal transport associated with double-ellipsoidal heat sources. Baeva et al.~\cite{baeva1997analysis} performed analysis of the temperature dynamics set by an elliptical cylindrical heat source (e.g. the electron beam) moving in a homogeneous plate, and obtained the temperature profiles as infinite series of the Mathieu-Hankel functions. Hou and Komanduri \cite{hou2000general} provided the general solutions in the integral form for the temperature profiles set in the sample from stationary/moving heat sources, including the elliptical ones. Wang et al.~\cite{wang2006numerical} performed numerical simulations of the laser welding, including double-ellipsoidal surface heat sources. Fachinotti et al.~\cite{fachinotti2011analytical} found analytical integral representations for the temperature field formed in a semi-infinite body by a moving double-ellipsoidal heat sources. Zhang et al.~\cite{zhang2011direct} employed the results \cite{nissim1980temperature} for a theoretical study of the impact that the elliptical beam parameters have on the direct laser patterning of self-assembled monolayers. Garcia-Garcia et al.~\cite{garcia2016simplified} employed the numerical approach to study the arc welding incorporating the double-ellipsoid model for the heat sources. Chiba et al.~\cite{chiba2017numerical} performed numerical analysis of temperature dynamics in the microelectronics sandwich structures associated with the glass-frit bonding by circular and elliptical laser beams. Mirkoohi et al.~\cite{mirkoohi2019heat} studied the selective laser melting with the aid of the integral temperature distributions obtained for semi-elliptical and double ellipsoidal moving heat sources. Models of ellipsoidal laser heat sources were used to study molten material dynamics in \cite{roehling2017modulating,abadi2021effect,sun2023systematic}. Ke et al.~\cite{ke2026thermal} employed the double ellipsoidal heat source model for thermo-mechanical analysis of welding deformations. An overview of the studies on temperature dynamics set by various heat sources can be found in \cite{nascimento2023literature}.

The obtained analytical solutions for the temperature profiles set in samples by elliptical heat sources were obtained either in the integral form, e.g. \cite{nissim1980temperature,moody1982temperature,hou2000general,fachinotti2011analytical}, or as infinite series of special functions, \cite{toshiyuki1982heat,baeva1997analysis}. While the latter approach admits computationally efficient approximations \cite{bremer2019algorithm,brimacombe2021computation}, the first approach involves slow and costly integral evaluations. In this contribution, we refine the analytical relations for the sample temperature found in \cite{stepanenko2024dimensional} by developing a model of gas thermal desorption driven by the sample irradiation with an elliptic Gaussian laser beam. One of the key findings of the study is the tractable form of the temperature profile represented in the form of rapidly converging functional series, which is then conveniently plugged into the gas desorption model. The remainder of the paper is organized as follows. In section \ref{sec:analytical}, we derive the analytical expression for the sample temperature and verify it against the rigorous integral relation and the numerical results of non-linear modeling. The possibility of using the developed thermal model for simulations of the tritium desorption is considered in Sec.~\ref{sec:tritium_desoprtion}. The conclusions of the study are summarized in section \ref{sec:conclusions}.

\section{Heat transport driven by elliptically shaped Gaussian laser beam\label{sec:analytical}}
\begin{figure}[ht!]
	\centering
	\includegraphics[scale=0.7]{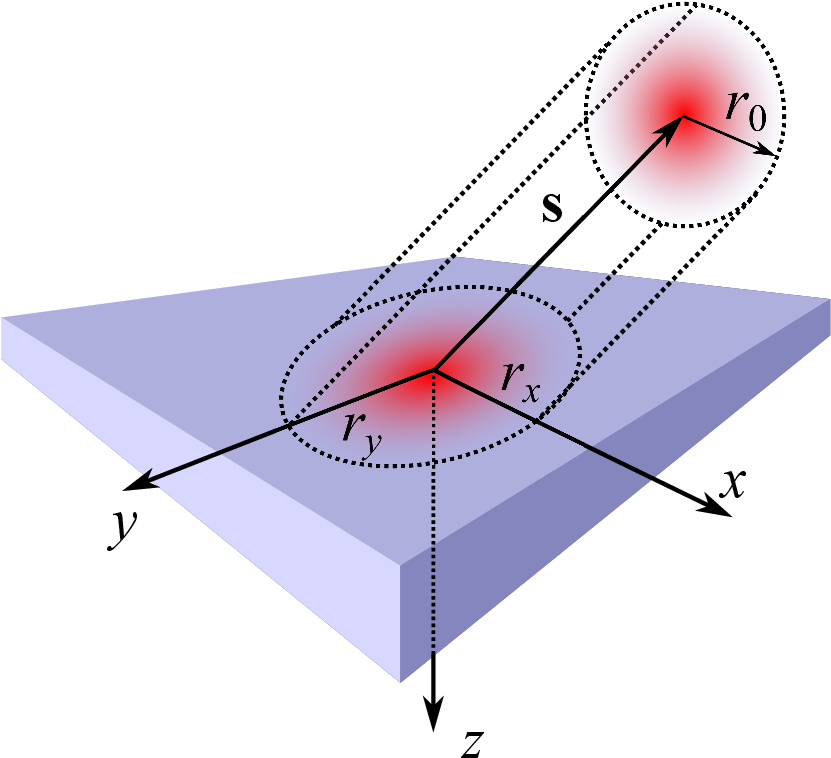}
	\caption{The sketch of the problem geometry. Details of obtaining the laser spot semi-axes $r_x$ and $r_y$ from the initial beam radius $r_0$ are given in Appendix A.\label{fig:geometry_sketch}}
\end{figure}

The analytical model of heat transfer in a solid employed for the study is similar to \cite{stepanenko2024dimensional}. We consider the semi-infinite planar sample. The sample characteristics, viz. the mass density, $\rho$, thermal conductivity, $\kappa$, and specific heat, $C_p$, are considered constant. The sample surface is irradiated by the laser beam with the following temporal profile and spatial distribution of the intensity in the spot,
\begin{eqnarray}
	I(x,y,t) = I_0 \exp\left(-\frac{x^2}{r_x^2}-\frac{y^2}{r_y^2}\right) H(t_p - t)\left(1-\Delta\frac{t}{t_p}\right) \equiv I_s(x,y)Y(t),
\end{eqnarray}
where $I_0$ is the beam intensity in the spot center, $t_p$ is the pulse duration, $r_x$ and $r_y$ are the profile semi-axes in the $x$ and $y$ directions, respectively, $\Delta$ is the pulse attenuation parameter, $H(t)$ is the Heaviside function, $I_s(x,y) = I_0\exp(-x^2/r_x^2-y^2/r_y^2)$ and $Y(t) = H(t_p - t)(1 - \Delta{t/t_p})$ are the spatial and temporal parts of $I$, respectively. In contrast to \cite{stepanenko2024dimensional}, we work in the Cartesian coordinate system, with the $z$ axis running normally to the surface, into the sample bulk, and the $x$ and $y$ axes are aligned along the ellipse semi-axes. Without loss of generality, we additionally assume that $r_y > r_x=r_0$, where $r_0$ is the initial beam radius. The sketch of the problem geometry is shown in Fig.~\ref{fig:geometry_sketch}. The semi-axes of the spatial profile, $r_x$ and $r_y$, can be obtained from $r_0$, as discussed in Appendix A.

The heat transfer in the sample is determined by thermal conductivity. On the sample surface, no phase transitions occur under the laser irradiation, and there are no energy sinks due to the convection or black-body radiation. To further simplify the problem, the energy source due to the laser heating is moved from the boundary into the sample bulk, similarly to \cite{stepanenko2024dimensional}. The resulting equations, governing the temperature evolution, become as follows,
\begin{eqnarray}
	\ddtp{T} = a^2 \left(\frac{\partial^2}{\partial{x}^2} + \frac{\partial^2}{\partial{y}^2} + \frac{\partial^2}{\partial{z}^2}\right)T + P, \label{eqn:T}
\end{eqnarray}
where $a^2 = \kappa/(\rho C_p)$ is the thermal diffusivity, 
\begin{equation}
	P(x,y,z,t) = P_0 \exp\left(-\frac{x^2}{r_x^2}-\frac{y^2}{r_y^2}\right) \delta(z) Y(t),
\end{equation}
$P_0 = 2I_0/(\rho C_p)$, and $\delta(z)$ is the Dirac $\delta$-function.

The boundary conditions imposed on $T$ are
\begin{eqnarray}
	\frac{\partial{T}}{\partial{z}}(x,y,0,t) = 0, \qquad \lim\limits_{x,y,z\rightarrow\infty}T(x,y,z,t) = T_0, \label{eqn:BC}
\end{eqnarray}
where $T_0$ is the thermostat temperature. The initial condition is
\begin{equation}
	T(x,y,z,0) = T_0. \label{eqn:IC}
\end{equation}

The general solution to Eqs.~(\ref{eqn:T}), (\ref{eqn:BC}), (\ref{eqn:IC}) can be represented as
\begin{equation}
	\frac{T(x,y,t)}{T_0} = 1 + J_0 \int_{-\infty}^{+\infty}d\mu f(\mu) \int_{\theta_{\min}}^{\theta_{\max}} d\theta \frac{h(\theta)}{1 + \theta^2} \exp\left[-\frac{\rho^2}{1 + \theta^2}\right], \label{expr:T_general}
\end{equation}
where $J_0 = I_0 r_x/(T_0 \kappa \sqrt{\pi})$, $h(\theta) = 1 - \Delta(\theta_{\max}^2 - \theta^2)/\theta_p^2$, $\theta_{\max} = 2a\sqrt{t}/r_x$, $\theta_p = 2a\sqrt{t}/r_x$, and $\rho^2 = x^2 + (y - \mu)^2$. The parameter $\theta_{\min} = 0$, if $t \leq t_p$, otherwise $\theta_{\min} = 2a\sqrt{t - t_p}/r_x$, and the function $f(\mu)$ is
given in Appendix B. Notice the absence of the dependence on the coordinate $z$, since we are looking for the temperature profile inside the layer, where the gas diffusion occurs. This region has the thickness on the order of the particle diffusion length, $l_D \sim \sqrt{Dt_p}$, where $D$ is the gas (tritium) diffusion coefficient. By estimating $D\sim10^{-3}\ \mathrm{cm^2/s}$ \cite{longhurst2008tmap7}, $t_p \sim 1$~ms \cite{zlobinski2011laser}, we find that $l_D \sim 10\ \mathrm{\mu{m}}$, which is by an order of magnitude smaller than the thermal diffusion length, $l_T \sim \sqrt{a^2 t_p}$, estimated for tungsten \cite{zlobinski2011laser}, $a^2 \sim 0.1\ \mathrm{cm^2/s}$, as $l_T \sim 0.1$~mm. As a result, the depth variation of $T$ is neglected in the further analysis.

The expression (\ref{expr:T_general}) can be simplified, as discussed in Appendix B, to arrive at the following final form of the relation for the sample temperature,
\begin{equation}
	\frac{T(x,y,t)}{T_0} = 1 + J_0 \exp\left(-\frac{x^2}{r_x^2} - \frac{y^2}{r_y^2}\right) \left[F(x,y,\theta_{\max}) - F(x,y,\theta_{\min})\right], \label{expr:T_polynomial}
\end{equation}
where
\begin{equation}
	F(x,y,\theta) = \sum_{n=0}^{\infty} R_n(x,y)\left[K_1 A_n(\theta) + K_2 B_n(\theta)\right],
\end{equation}
the coefficients $K_1 = 1 - \Delta\left(\theta_{\max}/\theta_p\right)^2$, $K_2 = \Delta/\theta_p^2$, and
\begin{eqnarray}
	A_n(\theta) = \frac{1}{n!} \int_{0}^{\theta} d\varphi \frac{\varphi^{2n}}{\left(1 + \varphi^2\right)^{n+1}},\\
	B_n(\theta) = \frac{1}{n!} \int_{0}^{\theta} d\varphi \frac{\varphi^{2n+2}}{\left(1 + \varphi^2\right)^{n+1}}.
\end{eqnarray}
The distance functions, $R_n(x,y)$, are given by
\begin{equation}
	R_n(x,y) = \sum_{m=0}^{n}\sum_{l=0}^{n-m} C_n^m C_{2n - 2m}^{2l} \frac{(2l-1)!!}{2^l} \xi^{2m} (k\eta)^{2n-2m-2l} e^{2l},
\end{equation}
where $\xi = x/r_x$, $\eta = y/r_y$ are the dimensionless coordinates of the point on the sample surface, $e=\sqrt{1-k^2}$ and $k = r_x/r_y$ are, respectively, the eccentricity and inverse aspect ratio of the laser beam spot. In the limit $k = 1, e=0$ ($r_x = r_y = r_0$, the circular beam shape), the factors $R_n(x,y)$ reduce to $R_n(x,y) = (\xi^2 + \eta^2)^n \equiv (r/r_0)^{2n}$, which is identical to the results \cite{stepanenko2024dimensional}.

The details behind the derivation of the general, (\ref{expr:T_general}), and approximate, (\ref{expr:T_polynomial}), representations for the sample temperature are given in Appendix B. Explicit expressions for the first few polynomials $A_n$, $B_n$, and $R_n$ are presented in Appendix C. Notice that, compared to, e.g.,  \cite{nissim1980temperature,moody1982temperature,toshiyuki1982heat,baeva1997analysis,hou2000general,fachinotti2011analytical}, the derived relation for the temperature distribution (\ref{expr:T_polynomial}) is expressed through \textit{elementary} functions, involves no integral evaluations, and can be easily evaluated up to the required order of accuracy.

To probe the validity of the obtained analytical relation (\ref{expr:T_polynomial}), we perform a test simulation of the sample temperature dynamics. For the test, we use $r_x = 0.5$~mm, $r_y = 1.5$~mm, set the beam intensity at the spot $I_0 = 850$~MW/m$^2$, pulse duration $t_p = 3$~ms, and the attenuation parameter $\Delta = 0$ (the rectangular pulse). For the sample material, we employ tungsten \cite{zlobinski2011laser}, $\rho = 19.079$~g/cm$^3$, $\kappa=118$~W/($\mathrm{m\cdot{}^\circ{C}}$), $C_p = 144$~J/($\mathrm{kg\cdot{}^\circ{C}}$).

The distribution of the sample temperature will be determined at the end of the heating pulse, when the temperature reaches its maximum and the approximation errors are the largest. For the function $F$ in (\ref{expr:T_polynomial}), we shall use six polynomials in the expansion. Properly resolving the elongation of the laser spot shape requires a larger number of polynomials in the approximating relation (\ref{expr:T_polynomial}), compared to the circular case of Ref.~\cite{stepanenko2024dimensional}, where taking two terms into account was sufficient to approximate $T$ with the error not exceeding $2-3\%$ in the center of the spot, compared to the exact profile. As will be demonstrated below, taking six terms in the expansion for the considered beam parameters is sufficient to recover the temperature profile in the spot center with the adequate accuracy.

\begin{figure}[ht!]
	\centering
	\includegraphics{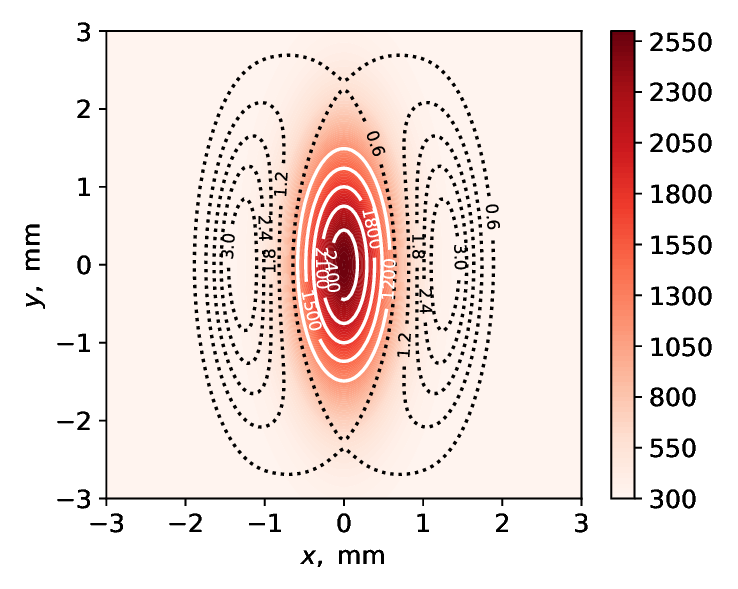}
	\caption{The sample temperature under irradiation by the elliptic laser beam with the semi-axes $r_x = 0.5$~mm, $r_y = 3$~mm at the pulse end, $t = t_p = 3$~ms. The solid lines represent the exact numerical solution to the heat equation (\ref{expr:T_general}). The dotted lines correspond to the relative error (in \%) between the approximate and exact solutions, $\delta = (T_{ex}-T_{appr})/T_{ex}\cdot100\%$. \label{fig:T_calculated}}
\end{figure}

Fig.~\ref{fig:T_calculated} demonstrates the temperature profiles for the sample surface, found by using the approximate solution (\ref{expr:T_polynomial}) and the general integral representation (\ref{expr:T_general}). As seen, profiles correspond closely to each other, with the relative error, $\delta = (T_{ex}-T_{appr})/T_{ex}$, not exceeding $0.6\%$ within the spot center, where the temperature takes the largest values and the desorption flux is maximal. At the distance on the order of $r_x$, the temperature deviations grow, however the gas desorption in these regions must be smaller compared to the spot center. The validity of this statement will be shown in a separate simulation of the tritium desorption provided in Sec.~\ref{sec:tritium_desoprtion}.

\begin{figure}[ht!]
	\centering
	\includegraphics{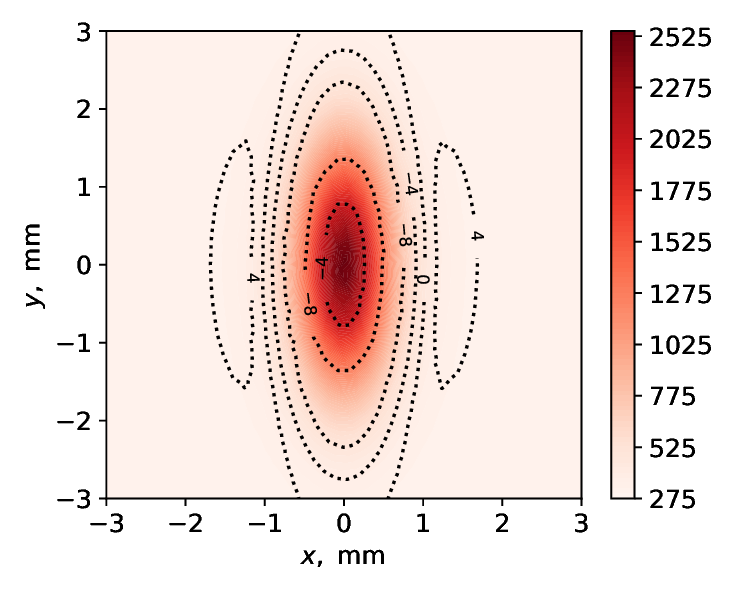}
	\caption{The sample temperature under irradiation by the elliptic laser beam with the semi-axes $r_x = 0.5$~mm, $r_y = 3$~mm at the pulse end, $t = t_p = 3$~ms, simulated by COMSOL Multiphysics. The dotted lines represent the relative deviation (in \%) of the simulated $T_{sim}$ profile from the approximate solution, $T_{appr}$, shown in  Fig.~\ref{fig:T_calculated}, $\delta = (T_{sim}-T_{appr})/T_{sim}\cdot100\%$.\label{fig:T_Comsol}}
\end{figure}

For comparison, we also present the temperature profile simulated with the COMSOL Multiphysics package \cite{comsol} under the similar irradiation conditions, Fig.~\ref{fig:T_Comsol}. The tungsten properties were defined with data by Ho et al. \cite{ho1972thermal}. For simulations, the heat transfer model was defined as follows. Modeling was carried out in the rectangular domain with the side lengths along $x$ and $y$ axes, $2L_x = 2L_y = 10$~mm, and the height along the $z$ direction, $H = 3$~mm. The modeling was run up to $t = 10$~ms. The boundary and initial conditions were
\begin{eqnarray}
	-\kappa\partial_z{T}(x,y,0,t) = I_s(x,y,t),\\
	\partial_{x}{T}(\pm L_x,y,z,t) = \partial_{y}{T}(x,\pm L_y,z,t) = \partial_z{T}(x,y,H,t) = 0,\\
	T(x,y,z,0) = T_0.
\end{eqnarray}

As seen in Fig.~\ref{fig:T_Comsol}, the approximate solution and the simulated temperature profile are very similar. The relative error does not exceed $4\%$ in the central area of the laser spot, which is comparable to the estimation errors found in Ref.~\cite{stepanenko2024dimensional}. At the spot edge, the linear model systematically overestimates the sample temperature  by more than limiting 5~\% \cite{stepanenko2024dimensional}, however in this area the desorption flux must be significantly smaller than in the spot center due to the strongly non-linear dependence of the desorption flux on the sample temperature. Therefore, (\ref{expr:T_polynomial}) can be used to close the thermal desorption equations for the analysis of the LID-QMS data. 

{We note that the validation of the derived thermal model (\ref{expr:T_polynomial}) was not possible due to the lack of the appropriate experimental data. Hence, only the verification of (\ref{expr:T_polynomial}) against the COMSOL Multiphysics simulations was performed.}

In addition to (\ref{expr:T_polynomial}), it is instructive to obtain a relation that would measure how taking the beam ellipticity into account influences on the temperature distribution compared to the circular beam case. As the target parameter, we shall use the temperature in the laser spot center determined at the end of the laser pulse, $T_m$, when deviations in the reconstructed temperature field would lead to the largest numerical errors in the gas desorption dynamics. For $T_m$, from (\ref{expr:T_polynomial}) we have $\theta_{\max} = \theta_{p}$, $\theta_{\min} = 0$, $F(x,y,0) = 0$ and
\begin{equation}
	\frac{T_m}{T_0} = 1 + J_0 \Biggl\{\left[K_1 A_0(\theta_{p}) + K_2 B_0(\theta_{p})\right] + \sum_{n=1}^{N_m}\frac{e^{2n}(2n-1)!!}{2^n}\left[K_1 A_n(\theta_{p}) + K_2 B_n(\theta_{p})\right]\Biggl\}, \label{eqn:Tm_elliptic}
\end{equation}
where $N_m$ is the approximation order, and, symbolically, $(-1)!!=1$. For the limiting case of the circular laser spot, $r_x = r_y = r_0$, we have $k = 1$, $e = 0$, and (\ref{eqn:Tm_elliptic}) reduces to
\begin{equation}
	\frac{T_m}{T_0} = 1 + J_0 \left[K_1 A_0(\theta_{p}) + K_2 B_0(\theta_{p})\right]. \label{eqn:Tm_round}
\end{equation}

By introducing (the indices \textit{e} and \textit{c} denote the elliptical and circular spot cases, respectively)
\begin{eqnarray}
	\eqalign{\Delta{T_m^e} &\equiv \frac{T_m^e}{T_0} - 1 \cr &= J_0^e \Biggl\{\left[K_1 A_0(\theta_{p}) + K_2 B_0(\theta_{p})\right] + \sum_{n=1}^{N_m}\frac{e^{2n}(2n-1)!!}{2^n}\left[K_1 A_n(\theta_{p}) + K_2 B_n(\theta_{p})\right]\Biggl\},}\\
	\Delta{T_m^c} = J_0^c \left[K_1 A_0(\theta_{p}) + K_2 B_0(\theta_{p})\right],
\end{eqnarray}
and recalling that
\begin{equation}
	\frac{J_0^e}{J_0^c} = \frac{r_x}{r_y} \equiv k,
\end{equation}
we find the required error measure
\begin{equation}
	\varepsilon_T = \frac{\Delta{T_m^e}}{\Delta{T_m^c}} - 1 \equiv \frac{T_m^e - T_m^c}{T_m^c - T_0} = (k-1)  + k\sum_{n=1}^{N_m}\frac{e^{2n}(2n-1)!!}{2^n} \delta_n, \label{expr:deltaT}
\end{equation}
where 
\begin{equation}
	\delta_n = \frac{K_1 A_n(\theta_{p}) + K_2 B_n(\theta_{p})}{K_1 A_0(\theta_{p}) + K_2 B_0(\theta_{p})}.
\end{equation}
The parameter $\varepsilon_T$ determines the relative exceedance of the maximal temperature in the center of the elliptic spot compared to $T_m$ in the center of the circular laser spot. 

The expression (\ref{expr:deltaT}) was derived by taking the total power deposited within the laser spot to be constant, i.e. $W_0 = I_0\pi r_xr_y = \const$. In case when the beam intensity in the spot center is constant between the elliptic and circular spot cases, $I_0 = \const$, (\ref{expr:deltaT}) is modified as
\begin{equation}
	\varepsilon_T = \sum_{n=1}^{N_m}\frac{e^{2n}(2n-1)!!}{2^n} \delta_n. \label{expr:deltaT_mod}
\end{equation}

The factors $\delta_n$ entering (\ref{expr:deltaT}), (\ref{expr:deltaT_mod}) can be estimated as follows. For the rectangular pulse, $\Delta = 0$, one has $K_1 = 1, K_2 = 0$, and the relations for $\delta_n$ reduce to
\begin{equation}
	\delta_n = \frac{A_n(\theta_{p})}{A_0(\theta_{p})}.
\end{equation}
For the triangular pulse, $\Delta = 1$, we have $K_1 = 0, K_2 = 1/\theta_{p}^2$, and
\begin{equation}
	\delta_n = \frac{B_n(\theta_{p})}{B_0(\theta_{p})}.
\end{equation}

As a crude estimate for $\varepsilon_T$, we can take $N_m = 1$ to arrive at
\begin{eqnarray}
	\varepsilon_T = k\left(1 + \frac{e^2}{2}\delta_1\right) - 1 \longleftrightarrow W_0 = \const,\\
	\varepsilon_T = \frac{e^2}{2}\delta_1 \longleftrightarrow I_0 = \const.
\end{eqnarray}
In the long-pulse limit, $\theta_{p} \gg 1$ (e.g., for tungsten, taking $t_p = 10$~ms, $a^2 = 4.3\times10^{-5}$~m$^2$/s, $r_0 = 0.5$~mm, one finds $\theta_{p} \approx 2.6$), we have (see Appendix C),
\begin{eqnarray}
	A_0(\theta_{p}) \approx 2A_1(\theta_{p}) \approx \arctan \theta_{p}, \\
	B_0(\theta_{p}) \approx B_1(\theta_{p}) \approx \theta_{p}.
\end{eqnarray}
In the short-pulse approximation, $\theta_p \ll 1$, we analogously find
\begin{eqnarray}
	A_0(\theta_{p}) \approx \theta_p, \qquad
	A_1(\theta_{p}) = B_0(\theta_{p}) \approx \frac{\theta_p^3}{3}, \qquad
	B_1(\theta_{p}) \approx \frac{\theta_p^5}{5}.
\end{eqnarray}

As a result, in case $W_0 = \const$, one has
\begin{eqnarray}
	\varepsilon_T &= k\left(1 + \frac{e^2}{4}\right) - 1, \qquad &\theta_p \gg 1,\ \Delta = 0,\\ 
	\varepsilon_T &= k\left(1 + \frac{e^2}{2}\right) - 1, \qquad &\theta_p \gg 1,\ \Delta = 1,\\
	\varepsilon_T &= k\left(1 + \frac{e^2\theta_p^2}{6}\right) - 1, \qquad &\theta_p \ll 1,\ \Delta = 0,\\ 
	\varepsilon_T &= k\left(1 + \frac{3e^2\theta_p^2}{10}\right) - 1, \qquad &\theta_p \ll 1,\ \Delta = 1.
\end{eqnarray}
In the large aspect-ratio limit, $k \ll 1$, and the above relations yield the two-fold error between the elliptic and circular spot profiles, $\varepsilon_T = -1$ (since the fixed power $W_0$ is redistributed across the area $\gg \pi r_0^2$, so that $T_m^e \rightarrow T_0$).

In case $I_0 = \const$, we find 
\begin{eqnarray}
	\varepsilon_T &= \frac{e^2}{4}, \qquad &\theta_p \gg 1,\ \Delta = 0,\\ 
	\varepsilon_T &= \frac{e^2}{2}, \qquad &\theta_p \gg 1,\ \Delta = 1,\\
	\varepsilon_T &= \frac{e^2\theta_p^2}{6}, \qquad &\theta_p \ll 1,\ \Delta = 0,\\ 
	\varepsilon_T &= \frac{3e^2\theta_p^2}{10}, \qquad &\theta_p \ll 1,\ \Delta = 1.
\end{eqnarray}
Interestingly, in this limit, the ratio $\varepsilon_T(\Delta = 1)/\varepsilon_T(\Delta=0)$ remains practically the same. For $\theta_p \ll 1$ one has $\varepsilon_T(\Delta = 1)/\varepsilon_T(\Delta=0) = 1.8$, whereas for $\theta_p \gg 1$ one finds $\varepsilon_T(\Delta = 1)/\varepsilon_T(\Delta=0) = 2$. 

For the trapezoidal pulses, the values of $\varepsilon_T$ fall in between the limiting values for the rectangular ($\Delta = 0$) and triangular ($\Delta = 1$) pulses. The marginal value $\varepsilon_T=0$ corresponds to $k \rightarrow 1$, $e \rightarrow 0$, i.e. when the ellipse transforms into the circle. 

\begin{figure}[ht!]
	\centering
	\includegraphics[width=.45\textwidth]{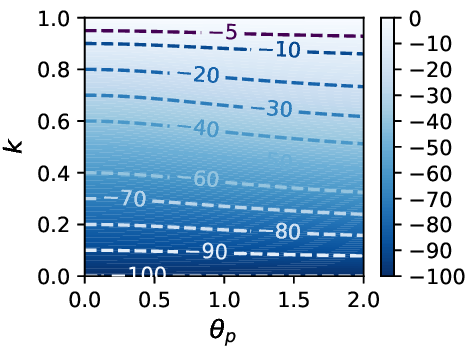}
	\includegraphics[width=.45\textwidth]{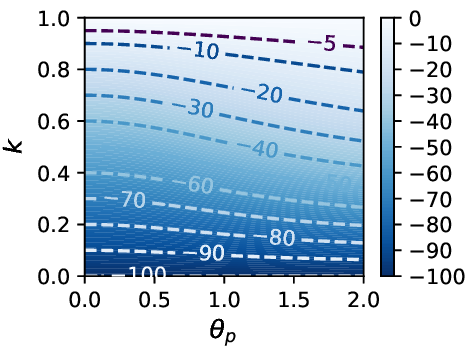}
	\caption{Dependence of $\varepsilon_T$ (in \%) on the beam inverse aspect ratio $k$ and normalized pulse duration $\theta_{p} = 2a\sqrt{t_p}/r_x \equiv 2a\sqrt{t_p}/r_0$ in the constant deposited power case, $W_0 = \const$. Left panel -- rectangular laser pulse, $\Delta = 0$, right panel -- triangular laser pulse, $\Delta = 1$.\label{fig:DeltaT_mod}}
\end{figure}

The distributions $\varepsilon_T(\theta_p, k)$ in the constant deposited power approximation, $W_0 = \const$, are shown in Figures~\ref{fig:DeltaT_mod} for the rectangular and triangular pulses. As seen, the increase of the laser spot aspect ratio results in the growth of $|\varepsilon_T|$ indicating the accumulation of approximation errors between the elliptic and circular spot cases. The 5~\% threshold error \cite{stepanenko2024dimensional} is reached for $k \leq 0.9-0.95$, which is equivalent to $r_y/r_x \geq (10/9)\div(20/19)$, i.e. even small variations in the spot shape ($\sim 5-10\%$) result in noticeable distortions in the temperature distribution within the spot area.

\begin{figure}[ht!]
	\centering
	\includegraphics[width=.45\textwidth]{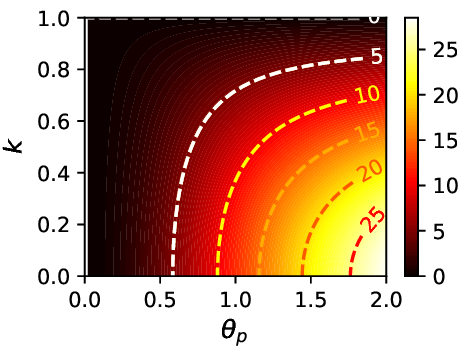}
	\includegraphics[width=.45\textwidth]{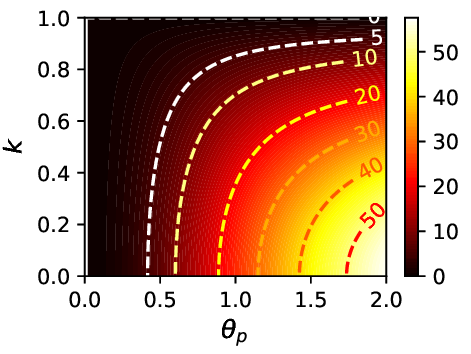}
	\caption{Dependence of $\varepsilon_T$ (in \%) on the beam inverse aspect ratio $k$ and normalized pulse duration $\theta_{p} = 2a\sqrt{t_p}/r_x \equiv 2a\sqrt{t_p}/r_0$ in the constant maximal spot intensity case, $I_0 = \const$. Left panel -- rectangular laser pulse, $\Delta = 0$, right panel -- triangular laser pulse, $\Delta = 1$.\label{fig:DeltaT}}
\end{figure}

In case $I_0 = \const$, the dependencies $\varepsilon_T(\theta_p, k)$ for the rectangular and triangular laser pulses are shown in Figures~\ref{fig:DeltaT}. As seen, the increase of the pulse duration leads to the growth of $\varepsilon_T$. For $\theta_{p} \le 0.5$, $\varepsilon_T$ remains below the 5\% limiting threshold \cite{stepanenko2024dimensional} in the whole range of the parameter $k$. For $\theta_p \ge 0.5$, the measure $\varepsilon_T > 5\%$, when $k < 0.8 \leftrightarrow r_y/r_x > 5/4$ (for the rectangular pulse) and when $k < 0.9 \leftrightarrow r_y/r_x > 10/9$ (for the triangular pulse). The shape distortion ($|r_y/r_x-1|\cdot100\%$) required to noticeably alter the temperature distribution within the spot reaches $\sim 10-25\%$.

Thus, when comparing the temperature distributions found for the elliptic and circular laser spots, the spot ellipticity must be already taken into account even for moderate elongations of the spot shape, when the large semi-axis of the ellipse exceeds the small semi-axis by approximately $5-10\%$. This result is valid in the constant deposited power approximation and is primarily due to the energy redistribution within the spot, leading to the reduction of the peak intensity. When comparing temperature distributions found for two laser spots having the same peak intensity, the beam shaping effects become important when the ellipse semi-axes differ by $\ge 10-25\%$. 

{The identified thresholds on the parameters of the beam incidence will vary depending on the analyzed material, condition of the sample surface, duration and intensity of the laser beam, and other sample and beam characteristics. Nevertheless, the qualitative conclusion that for the fixed $W_0$ the condition for approximating the elliptical beam with the circular one is more stringent than in case $I_0 = \const$ will still remain valid and should be taken into account when analyzing experimental data.}

{We emphasize that both examined approximations, $W_0 = \const$ and $I_0 = \const$, should be considered as the limiting cases, which can be used to assess the applicability limits of the circular beam approximation for interpreting the LID-QMS data. These estimates demonstrate how uncertainties in the beam profile and intensity distribution along the sample surface translate into variations of the tritium desorption dynamics, depending on the available experimental apparatus.}

\section{3D tritium desorption dynamics\label{sec:tritium_desoprtion}}
To study dynamics of the tritium removal from the tungsten sample, we now couple the developed thermal model of heat transport, driven by the sample irradiation with the elliptic laser beam, to the tritium desorption model. The latter model assumes the tritium release from traps into the solution, followed by the atom diffusion to the sample surface and desorption into the gas phase as a result of the molecular recombination. The model, therefore, describes the spatial and temporal evolution of the tritium concentration in the solution, $u$, and in the traps, $y$,
\begin{eqnarray}
	\ddtp{u} = \nabla\cdot\left[D_0\exp\left(-\frac{E_d}{kT}\right)\nabla{u}\right] - \ddtp{y},\label{eqn:u}\\
	\ddtp{y} = \nu\exp\left(-\frac{E_d}{kT}\right) \left[\left(y_{m} - y\right)\frac{u}{u_m} 
	- \exp\left(-\frac{E_{b}}{kT}\right) y\right]. \label{eqn:y}
\end{eqnarray}
In these equations, $u_m$ is the concentration of the tungsten atoms in the sample, $D_0$, $E_d$ are, correspondingly, the diffusion pre-exponent and activation energy, $y_{m}$, $E_{b}$ are the maximum concentration and binding energy for gas particles in traps, $\nu=10^{13}\ \mathrm{s^{-1}}$ is the Debye frequency. The sample temperature $T$ is defined with (\ref{expr:T_polynomial}), taking six polynomials in the expansion for the approximant $F$. Equations are solved in the cylindrical coordinate system, $(r,\varphi,z)$.

Eqs.~(\ref{eqn:u}), (\ref{eqn:y}) are closed with the following initial and boundary conditions. In the initial moment of time, we assume that all gas atoms are contained within traps,
\begin{equation}
	u(r,\varphi,z,0) = 0, \qquad y(r,\varphi,z,0) = y_{m}.
\end{equation}
The boundary conditions are written in the limit of the instantaneous atom recombination at the sample surface \cite{kulagin2023effect}, by assuming that there are no particle fluxes at the domain boundaries in the sample bulk (i.e. at the sufficiently large distance from the laser spot),
\begin{equation}
	u(r,\varphi,0,t) = 0, \qquad \partial_r u(0,\varphi,z,t) = \partial_r u(R,\varphi,z,t) = \partial_z u(r,\varphi,L,t) = 0. \label{eqn:BC_desorption}
\end{equation}
In these relations, $R$ and $L$ are the radial and axial scales of the domain, respectively. 

{Equations (\ref{eqn:u})-(\ref{eqn:BC_desorption}) represent the the minimal working model of tritium desorption, which can be used to assess the impact of a new thermal transport model (\ref{expr:T_polynomial}). The rigorous desorption analysis must necessarily take into the presence of multiple type traps, their inhomogeneous distribution across the sample, balance of particle fluxes at the sample surface. Such a step goes beyond the scope of the present study.}

{We note that following the heat transfer problem the model (\ref{eqn:u})-(\ref{eqn:BC_desorption}) could also  be treated analytically. However, due to non-linearities entering (\ref{eqn:y}) and strong transients arising within the sample during the heating cycle, such a study will not be carried out in this manuscript and will be addressed elsewhere.}

For simulations, we set the following laser beam parameters: $I_0 = 850$~MW/m$^2$, $r_x = 0.5$~mm, $r_y = 1.5$~mm, $t_p = 3$~ms, $\Delta = 0$. Thus, the beam inverse aspect ratio and the eccentricity are $k=1/3$ and $e=0.943$, respectively. The sample desorption parameters were $L = 10\ \mathrm{\mu{m}}$, $R=6r_x = 3$~mm, $D_0 = 2\times10^{-3}$~cm$^2$/s, $E_d = 0.39$~eV, $u_m = 6.31\times10^{22}$~part./cm$^3$, $y_{m}/u_m = 10^{-2}$, $E_{b} = 1.5$~eV. The tungsten material properties were $\rho = 19.079$~g/cm$^3$, $\kappa=118$~W/($\mathrm{m\cdot{}^\circ{C}}$), $C_p = 144$~J/($\mathrm{kg\cdot{}^\circ{C}}$).

Simulations were carried out by using the numerical code written with the aid of the BOUT++ library \cite{dudson2009bout++,dudson2015bout++}. Since the sample thickness $L$ is much smaller than the beam radius $r_x$, the radial and azimuthal derivatives can be omitted in the del operators entering (\ref{eqn:u}), (\ref{eqn:y}). The remaining axial derivatives were discretized with the fourth-order accurate central scheme. Time-stepping was performed with the Backward Differentiation Formulas (BDF) scheme of the variable order up to 100 diffusion timescales, $t_s = 100t_D \equiv 100 (L^2/D_0)=50$~ms. The simulation grid had the resolution $N_r \times N_\varphi \times N_z = 64\times64\times64$.

\begin{figure}[ht!]
	\centering
	\includegraphics[scale=.75]{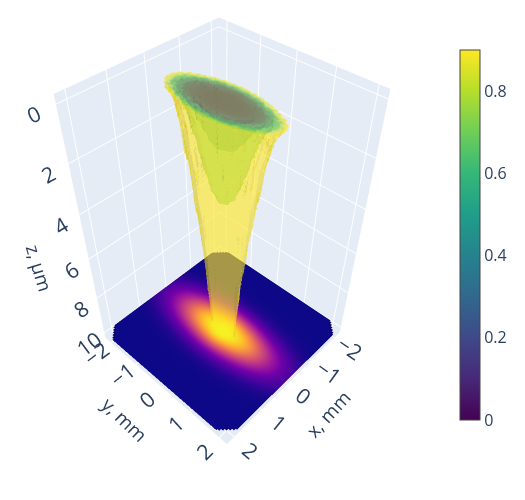}
	\caption{Normalized distributions of the tritium concentration in the traps, $\hat{y} = y/y_{m}$. The heat map at the bottom qualitatively shows the temperature distribution at end of the laser pulse, $t=t_p=3$~ms.\label{fig:y_profiles}}
\end{figure}

The distribution of tritium in traps found at the simulation end is demonstrated in Fig.~\ref{fig:y_profiles}. As seen, the particles desorb from the region with the elliptical symmetry conforming with the shape of the laser spot at the sample surface.

\begin{figure}[ht!]
	\centering
	\includegraphics[scale=1.]{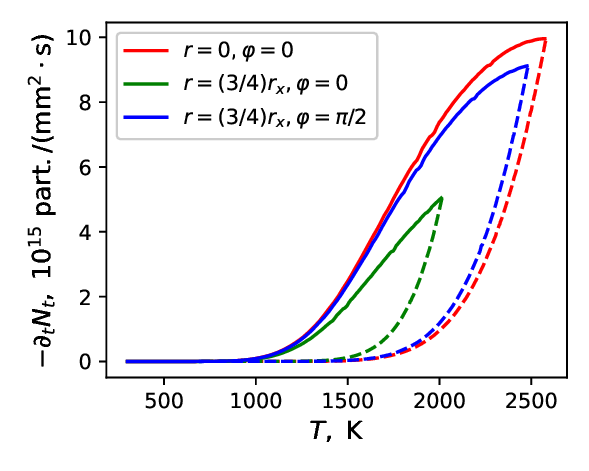}
	\caption{Tritium desorption rates along the sample surface, $\partial_t{N_t}$, vs. the local sample temperature $T$, $-\partial_t{N_t}[r,\varphi, T(t)] = -\partial_t\left[\int_0^{L}{dz}(y + u)(r,\varphi, z, t)\right]$. The solid/dashed curves correspond to the part of the desorption cycle associated with the laser heating/sample cooling. The red curve corresponds to the spot center, the green (blue) curve -- to the spot along the $x$ ($y$) axis.\label{fig:TPD}}
\end{figure}

{The local desorption rates (the TPD curves) found in three different spots along the sample surface, viz. in the spot center, $r = 0, \varphi = 0$, and along the $x$ and $y$ semi-axes of the temperature profile, $r = (3/4)r_x, \varphi = 0$ and $r = (3/4)r_x, \varphi = \pi/2$, respectively, are demonstrated in Fig.~\ref{fig:TPD}. As expected, with the increase of the distance from the spot the desorption rates become smaller as the local temperature decreases. Along the $x$ axis ($r = (3/4)r_x, \varphi = 0$) the temperature decrease is stronger than along the $y$ axis ($r = (3/4)r_x, \varphi = \pi/2$), resulting in a larger reduction of the tritium desorption rate. All curves shown in Fig.~\ref{fig:TPD} demonstrate hysteresis associated with different thermal transport dynamics during the laser heating and sample cooling cycles.}

\begin{figure}[ht!]
	\centering
	\includegraphics[scale=1.0]{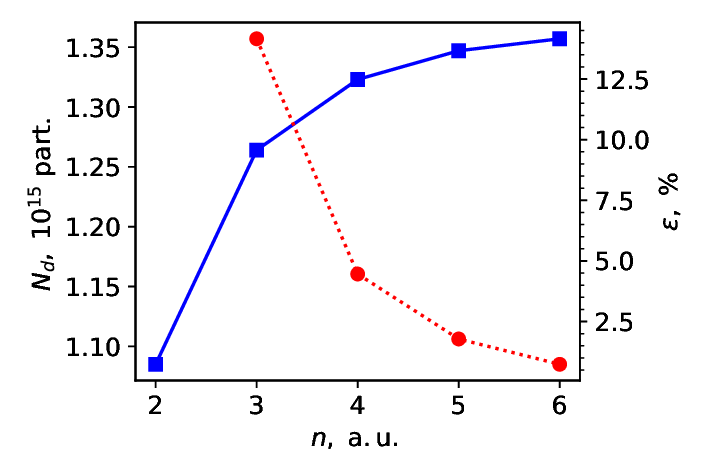}
	\caption{Dependence of the total desorbed tritium atoms, $N_d$, and the relative approximation error, $\varepsilon = [N_d(n)/N_d(n-1)-1]\cdot100\%$ on the approximation order, $n$, for the sample temperature $T$ (\ref{expr:T_polynomial}).\label{fig:desorbed_y_order}}
\end{figure}

Fig.~\ref{fig:desorbed_y_order} demonstrates the dependence of the total number of tritium particles desorbed from the sample at the pulse end, $N_d$,
\begin{equation}
	N_d(t) = \int_{0}^{R}rdr\int_{0}^{2\pi}d\varphi\int_{0}^{L}dz \left[y_m - y(r,\varphi,z,t\rightarrow\infty)\right],
\end{equation}
on the order of approximation, $n$, of the temperature profile (\ref{expr:T_polynomial}). As seen, for $n=6$ the relative error in determining $N_d$ becomes less than $1\%$, effectively indicating that using higher-order approximations, $n \geq 7$, for analysis of the LID-QMS data (at least, for beam aspect ratios $\leq 3$) becomes redundant.

Apart from testing the tritium desorption dynamics, the tritium removal model was also used to assess the applicability limits for the circular beam approximation for rapid analysis of LID-QMS data, under which the elliptic distortions of the spot shape related to non-normal incidence of the laser beam on the sample surface can be neglected. For simulations, the previously shown sample and laser source parameters were used, except for the pulse attenuation parameter and the spot minor and major semi-axes, which were defined as follows $\Delta = \{0, 1\}$, $r_x = r_0 = 1$~mm, $r_y = \{1, 1.05, 1.1, 1.2, 1.3, 1.4, 1.5, 2, 2.5, 3\}$~mm, respectively. For the considered parameters, the normalized pulse duration is $\theta_p \approx 0.7 \sim 1$, which falls in between the above limiting cases of short/long pulses. The modeling was performed in the constant deposited power / spot peak intensity approximations. The obtained data are summarized in Table~\ref{tab:Nd_vs_shape}. As expected, in the constant deposited power case, $W_0 = \const$, the increase of the spot aspect ratio leads to the reduction of the peak laser intensity in the spot center, compared to the circular spot case, notably reducing the desorption flux and the amount of desorbed tritium. Even moderate reduction of $T_m$ within 20\% range results in the overestimation of $N_d$ by 2-5 times within the circular beam approximation, depending on the attenuation parameter $\Delta$. For a strongly elongated spot, $r_y/r_x = 3$, the relative differences between values of $N_d$ in the elliptic and circular beam cases reach 2-3 orders of magnitude. The main driver for such strong variations in the resulting quantities is the power redistribution within the spot leading to the reduction of the peak power deposited on the sample surface in the spot center by $r_x/r_y$ times. On the contrary, in the constant peak intensity approximation, $I_0 = \const$, the amount of desorbed tritium in the circular beam case becomes underestimated, compared to the elliptic beam limit. The increase of the spot aspect ratio to $r_y/r_x = 3$ leads to the almost three/four-fold error in the magnitude of $N_d$, which is, however, much smaller than the orders-of-magnitude errors in case $W_0 = \const$. Notice that in all cases shown in Table~\ref{tab:Nd_vs_shape} uncertainties in the temperature distribution convert in at least an order of magnitude larger relative errors in $N_d$.

These results indicate that the desorption flux is highly sensitive to the temperature in the spot center (where most particles desorb from the sample), herein uncertainties in the spot shape become less important, when the peak intensity in the laser spot center is known. In cases, when exact values of $I_0$ in the spot center are unknown and are reconstructed from the initial beam intensity, even small variations in the spot shape can result in unacceptably large errors potentially reaching, in extreme cases, 2-3 orders of magnitude for the laser spot aspect ratios $r_y/r_x \ge 2-3$. Thus, to obtain a reliable rapid estimate for $N_d$, the elliptic beam in the tritium desorption model can be replaced with the circular one, provided the value of $I_0$ is a known parameter. In this limit, for the cases considered in the study the relative error in $N_d$ will remain $\le 50\%$, if the beam/spot aspect ratio remains reasonably small, $r_y/r_x \le 2.0$. In terms of the beam incidence angle $\alpha$ (see Appendix A), this condition corresponds to $\alpha \le \alpha_c \approx 60^\circ$, herein the critical angle $\alpha_c$ may vary in both directions, depending on the beam intensity, tritium deposition depth, material properties. In cases, when $I_0$ remains an unknown quantity, the application of the circular beam approximation remains valid for spot aspect ratios $r_y/r_x \le 1.2\ (1.1)$, when $\Delta=0\ (1)$. This corresponds to the beam incidence angles $\alpha < \alpha_c \approx 34^\circ\ (25^\circ)$ [$\Delta=0\ (1)$]. For beams falling on the sample surface at grazing angles, $\alpha > \alpha_c$, the circular spot approximation becomes inapplicable for coupling to the tritium desorption model regardless of the employed limit.
\begin{table}[ht!]
	\caption{\label{tab:Nd_vs_shape}The dependence of $N_d$ on the elliptic spot aspect ratio in the constant deposited power and peak spot intensity approximations. The relative errors $\varepsilon_N$ are evaluated with respect to the corresponding elliptic beam cases, $\varepsilon_N=[N_d(1)/N_d(k) - 1]\cdot100\%$. The relative error $\varepsilon_T^* = (\Delta{T_m^{c}}/\Delta{T_m^{e}} - 1)\cdot100\% \equiv -[\varepsilon_T/(1+\varepsilon_T)]\cdot100\%$. The measures $\varepsilon_N$ and $\varepsilon_T^*$ define the approximation errors in $N_d$ and $T_m$, respectively, obtained if the elliptic beam is replaced with the circular one.}
	\begin{indented}
		\item[]
		\begin{tabular}{@{}c|ccc|ccc}
			\br
			\multirow{2}{.75cm}{$r_y/r_x$} & \multicolumn{6}{c}{$W_0 = \const$}\\
			& $N_d, \Delta = 0$ & $\varepsilon_N,\ \%$ & $\varepsilon_T^*,\ \%$ & $N_d, \Delta = 1$ & $\varepsilon_N,\ \%$ & $\varepsilon_T^*,\ \%$\\
			\mr
			1	&	
			$2.35\times10^{15}$&$0$&$0$&
			$1.05\times10^{14}$&$0$&$0$\\
			1.05 &
			$2.08\times10^{15}$&$13.0$&$3.7$&
			$7.95\times10^{13}$&$32.1$&$2.6$\\
			1.1 &
			$1.88\times10^{15}$&$25.0$&$7.5$&
			$6.05\times10^{13}$&$73.6$&$5.3$\\
			1.2 &
			$1.40\times10^{15}$&$67.9$&$15.0$&
			$3.56\times10^{13}$&$194.9$&$10.8$\\
			1.3 &
			$1.13\times10^{15}$&$108.0$&$22.6$&
			$2.15\times10^{13}$&$388.4$&$16.4$\\
			1.4 &
			$8.68\times10^{14}$&$170.7$&$30.2$&
			$1.33\times10^{13}$&$689.5$&$22.1$\\
			1.5	&	
			$6.70\times10^{14}$&$250.7$&$37.9$&	
			$8.42\times10^{12}$&$1147.0$&$28.0$\\
			2	&	
			$1.72\times10^{14}$&$1266.3$&$77.0$&	
			$1.17\times10^{12}$&$8874.4$&$58.8$\\
			2.5	&	
			$4.74\times10^{13}$&$4857.8$&$116.9$&	
			$3.37\times10^{11}$&$31057.3$&$91.3$\\
			3	&	
			$1.43\times10^{13}$&$16333.6$&$157.3$&	
			$1.73\times10^{11}$&$60593.6$&$124.8$\\
			\br
		\end{tabular}
		\item[]
		\begin{tabular}{@{}c|ccc|ccc}
			\br
			\multirow{2}{.75cm}{$r_y/r_x$} & \multicolumn{6}{c}{$I_0 = \const$}\\
			& $N_d, \Delta = 0$ & $\varepsilon_N,\ \%$ & $\varepsilon_T^*,\ \%$ & $N_d, \Delta = 1$ & $\varepsilon_N,\ \%$ & $\varepsilon_T^*,\ \%$\\
			\mr
			1	&	
			$2.35\times10^{15}$&$0$&$0$&	
			$1.05\times10^{14}$&$0$&$0$\\
			1.05 &
			$2.45\times10^{15}$&$-4.1$&$-1.2$&
			$1.12\times10^{14}$&$-6.3$&$-2.3$\\
			1.1 &
			$2.60\times10^{15}$&$-9.6$&$-2.3$&
			$1.21\times10^{14}$&$-13.2$&$-4.3$\\
			1.2 &
			$2.94\times10^{15}$&$-20.1$&$-4.2$&
			$1.35\times10^{14}$&$-22.2$&$-7.7$\\
			1.3 &
			$3.34\times10^{15}$&$-29.6$&$-5.7$&
			$1.50\times10^{14}$&$-30.0$&$-10.5$\\
			1.4 &
			$3.49\times10^{15}$&$-32.7$&$-7.0$&
			$1.65\times10^{14}$&$-36.4$&$-12.8$\\
			1.5	&	
			$3.81\times10^{15}$&$-38.3$&$-8.1$&
			$1.80\times10^{14}$&$-41.7$&$-14.7$\\
			2	&	
			$5.28\times10^{15}$&$-55.5$&$-11.5$&	
			$2.51\times10^{14}$&$-58.2$&$-20.6$\\
			2.5	&	
			$6.67\times10^{15}$&$-64.8$&$-13.3$&	
			$3.22\times10^{14}$&$-67.4$&$-23.5$\\
			3	&	
			$8.20\times10^{15}$&$-71.3$&$-14.2$&	
			$3.90\times10^{14}$&$-73.1$&$-25.1$\\
			\br
		\end{tabular}
	\end{indented}
\end{table}

\section{Conclusions\label{sec:conclusions}}
In this contribution, we have analyzed the tritium removal from a solid sample driven by irradiation with a non-circular, elliptically shaped Gaussian laser beam. We have obtained the new analytically tractable expression describing the temporal and spatial evolution of the sample temperature, thus generalizing the results \cite{stepanenko2024dimensional} to elliptically-shaped laser spots/beams. The derived relation was verified against the rigorous integral representation for the sample temperature and the numerical simulations within the COMSOL Multiphysics package. Both tests yielded qualitatively and quantitatively close results. The six-term approximation produced the temperature profile that deviated from the rigorous analytical one by no more than 0.6\% within the central spot area at the end of the laser pulse. Compared to the results of non-linear numerical simulations, the approximate form of $T$ led to the temperature distribution that deviated from the modeling one by no more than 4\% in the spot center at the end of the laser pulse complying with the restrictions on the temperature approximation accuracy discussed in \cite{stepanenko2024dimensional}. The newly found expression for the sample temperature was supplied with the measure $\varepsilon_T$ that can be used to assess deviations in the temperature distributions driven by the elliptical distortions of the initially circular beam at the sample surface.

The derived formula for the sample temperature was coupled to the numerical tritium desorption model, which was further examined in a test case of tritium removal from a tungsten sample, irradiated by a laser beam with the aspect ratio $r_y/r_x=3$. The spatial distribution of tritium in traps, obtained at the end of the sample heating cycle, qualitatively complies with the elliptical shape of the laser spot on the sample surface. The comparison of the total numbers of particles desorbed from the sample, $N_d$, found by using various orders of approximation for the sample temperature, indicated that using the sixth-order approximation for $T$ is sufficient to effectively resolve the tritium desorption dynamics from a tungsten sample under irradiation by a laser beam with the aspect ratios $\leq 3$. 

The derived relations were used to assess the applicability limits for the circular beam approximation, when the elliptically-shaped beam/spot is replaced with the circular counter-part, producing a less accurate, yet computationally more rapid model for assessment of the total tritium desorbed from the sample. The comparison of the total amounts of desorbed tritium in the constant deposited power and peak spot intensity approximations demonstrated that $N_d$ is highly sensitive to variations in the peak spot intensity. Even moderate, within 20\% range, variations in the spot shape can lead up to a two/three-fold overestimation of $N_d$ based on the circular laser spot approximation, provided the peak intensity is unknown and is reconstructed at the sample surface from the initial beam intensity distribution. This result is a direct consequence of the power re-distribution within the spot when the beam aspect ratio increases. If the beam intensity is a known quantity, uncertainties in the spot shape produce smaller deviations in $N_d$, with the circular spot model underestimating the total amount of desorbed tritium. Under the considered laser source and material parameters, if $I_0$ is well-defined in the spot center, the $\le 50\%$ uncertainty in the magnitude of $N_d$ is achieved for the spot aspect ratios $<2$, which are accessible at the sub-critical beam incidence angles $\le \alpha_c \approx 60^\circ$. In case only the power deposited on the sample surface is known, the critical aspect ratios and incidence angles reduce to $r_y/r_x \le 1.1\div1.2$, $\alpha_c \approx 25^\circ\div34^\circ$, for the beam attenuation parameters $\Delta = 1\div0$. For beams irradiating the sample surface at grazing angles ($> \alpha_c$), using both the constant deposited power and peak intensity limits for the circular-beam approximation would yield unacceptably high errors in the values of $N_d$.

{The determined thresholds will naturally vary depending on variations in the sample and laser parameters. Nevertheless, the qualitative conclusion that the limits on the beam incidence angle are more tight in case only the total pulse energy is known will still hold and should be taken into account when analyzing experimental data. The accurate quantitative assessment of the measurement thresholds must be performed separately for each specific problem at hand.} 

\section*{ACKNOWLEDGMENTS}
The work was supported by the Ministry of Science and Higher Education of the Russian Federation (project No. FSWU-2024-0001).

\section*{AUTHOR DECLARATIONS}
\subsection*{Conflict of Interest}
The authors have no conflicts to disclose.

\section*{DATA AVAILABILITY}
The raw data can be obtained on request from the corresponding author.

\section*{Appendix A. Recovering semi-axes of the elliptic profile $I_s$ from the initial beam radius}	
\renewcommand{\theequation}{A.\arabic{equation}}
\renewcommand{\thefigure}{A.\arabic{figure}}
\setcounter{equation}{0}
\setcounter{figure}{0}

\begin{figure}[ht!]
	\centering
	\includegraphics[scale=0.7]{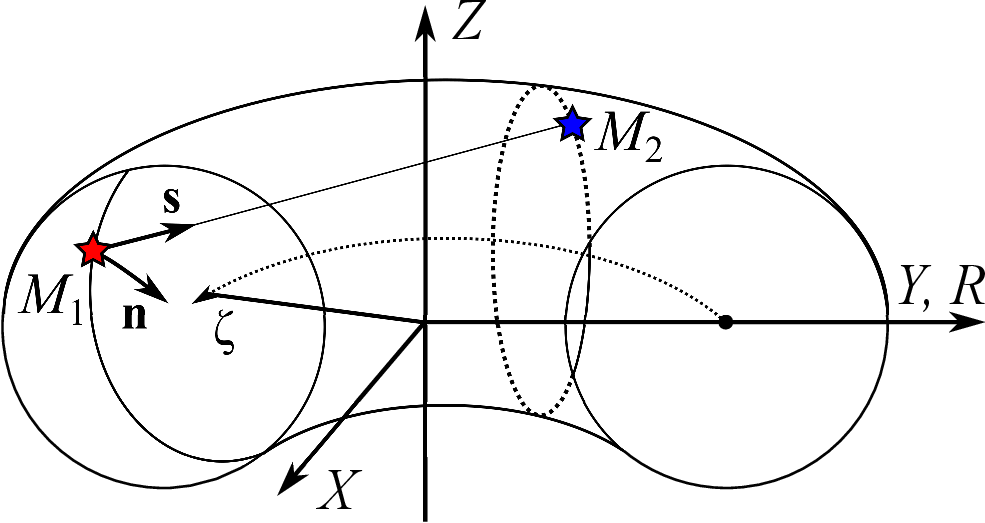}
	\caption{The sketch of the problem geometry. $M_1$ is the analysis point, and $M_2$ is the entry point for the laser beam. Note that $M_1$ is directly visible from $M_2$.\label{fig:tokamak_sketch}}
\end{figure}

In this Appendix, we shall derive relations coupling the semi-axes $r_x, r_y$ of the elliptic beam profile to the radius of the circular beam, $r_0$, initially produced by the laser light source. We shall assume that the analysis spot on the sample surface, $M_1$, and the laser beam entry point, $M_2$, are located arbitrarily inside the vacuum chamber of the tokamak, i.e. these points do not coincide and, in general case, do not lie in the same poloidal plane. The only requirement is that $M_1$ lies in the line-of-sight of $M_2$, Fig.~\ref{fig:tokamak_sketch}.

\begin{figure}[ht!]
	\centering
	\includegraphics[scale=0.7]{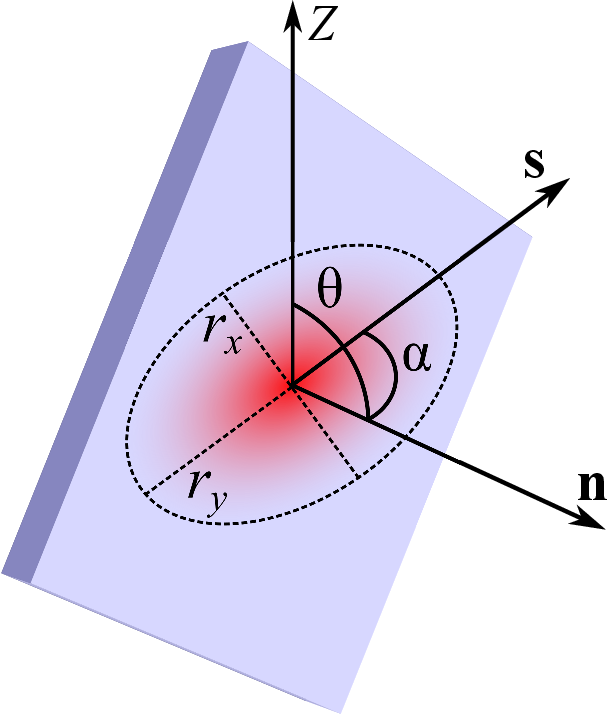}
	\caption{The orientation of the local spherical coordinate system, bind to the analyzed surface in the point $M_1$, to the global Cartesian/toroidal reference frame.\label{fig:3D_surface_sketch}}
\end{figure}

In the cylindrical coordinate system, the initial intensity distribution within the laser beam emitted from the source is
\begin{equation}
	I(r) = I_0\exp\left(-\frac{r^2}{r_0^2}\right).
\end{equation}
In the point $M_1$, the circular beam will have the elliptic shape due to the projection onto the oblique plane, located at the angle $\alpha$ with respect to the beam direction, Fig.~\ref{fig:3D_surface_sketch},
\begin{equation}
	I_s(x,y) = I'_0\exp\left(-\frac{x^2}{r_x^2}-\frac{y^2}{r_y^2}\right),
\end{equation}
where $x,y,z$ are the local Cartesian coordinates, bind to the sample surface in the point $M_1$, and $I'_0 \equiv I_0 [r_0^2/(r_xr_y)]$ is the beam intensity within the spot on the sample surface. The semi-axes, $r_x, r_y$, are given by
\begin{equation}
	r_x = r_0, \qquad r_y = r_0 \sec\alpha.
\end{equation}
In writing of these relations, we recall that the local axes $x$ and $y$ are aligned so as to ensure that $r_x < r_y$.

To find the angle $\alpha$, we construct the line segment connecting $M_1$ and $M_2$, Fig.~\ref{fig:tokamak_sketch}. Then we can define $\alpha$ as
\begin{equation}
	\cos\alpha = \mathbf{n}\cdot\mathbf{s},
\end{equation}
where $\mathbf{n}$ is the unit vector, normal to the sample surface in the point $M_1$, and $\mathbf{s}$ is the unit vector directed along the segment $M_1M_2$, Figs.~\ref{fig:tokamak_sketch}, \ref{fig:3D_surface_sketch}.

We shall also use the global Cartesian coordinate system $XYZ$ and the auxiliary toroidal reference frame, with the origins located in the same point on the machine symmetry axis, as shown in Fig.~\ref{fig:tokamak_sketch}. In the toroidal coordinates, the point position will be characterized by the $R, Z$ coordinates and also the toroidal angle $\zeta$.

The points $M_1$ and $M_2$ are defined in the toroidal coordinates as $M_1=(R_1,\zeta_1,Z_1)$, $M_2=(R_2,\zeta_2,Z_2)$. Then in the global Cartesian reference frame the vector $\mathbf{s}$ is uniquely defined as
\begin{equation}
	\mathbf{s}=\left(\frac{R_2}{L}\cos\zeta_2-\frac{R_1}{L}\cos\zeta_1; \frac{R_2}{L}\sin\zeta_2-\frac{R_1}{L}\sin\zeta_1; \frac{Z_2 - Z_1}{L}\right),
\end{equation}
where $L$ is the length of the line segment $M_1M_2$,
\begin{equation}
	L = \left[R_1^2 + R_2^2 - 2R_1R_2\cos\psi + \left(Z_2-Z_1\right)^2\right]^{1/2},
\end{equation}
where $\psi=\zeta_2-\zeta_1$ is the toroidal angle of the point $M_2$ with respect to $M_1$.

The normal vector $\mathbf{n}$ can be also defined in the global Cartesian reference frame by using the local spherical coordinates, tied to the analyzed sample surface in the point $M_1$. By aligning the polar axis along the $Z$ axis of the global coordinate system, we find in the $XYZ$ frame, Fig.~\ref{fig:3D_surface_sketch},
\begin{equation}
	\mathbf{n}=\left(\cos\zeta_1\sin\theta; \sin\zeta_1\sin\theta; \cos\theta\right),
\end{equation}
where $\theta$ is the angle between the axis $Z$ and the vector $\mathbf{n}$.

Combining expressions for the vectors $\mathbf{n}$ and $\mathbf{s}$ together, we find,
\begin{equation}
	r_y = \frac{r_0}{\mathbf{n}\cdot\mathbf{s}},
\end{equation}
where
\begin{equation}
	\mathbf{n}\cdot\mathbf{s} = \frac{\left(R_2\cos\psi-R_1\right)\sin\theta+\left(Z_2-Z_1\right)\cos\theta}{\left[R_1^2 + R_2^2 - 2R_1R_2\cos\psi + \left(Z_2-Z_1\right)^2\right]^{1/2}}. \label{expr:n_dot_s}
\end{equation}
The derived expression for $r_y$ is thus fully defined with the toroidal coordinates of the analysis and beam entry points.

\begin{figure}[ht!]
	\centering
	\includegraphics[scale=0.7]{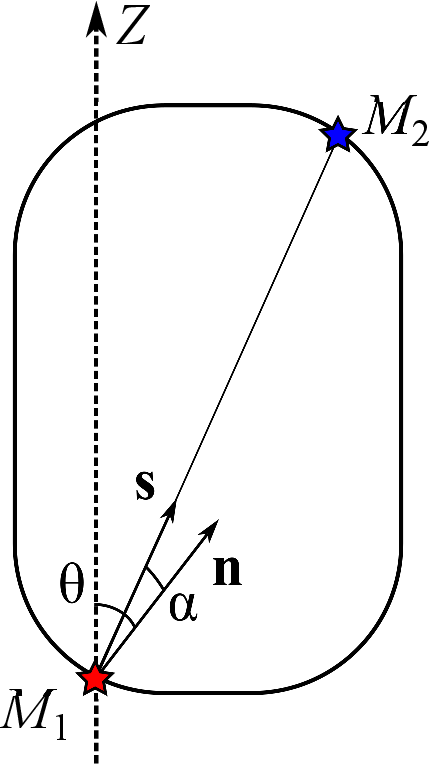}
	\caption{Geometry of the beam path in case when the points $M_1$ and $M_2$ lie in the same poloidal plane.\label{fig:2D_tokamak_sketch}}
\end{figure}

It is useful to derived a reduced expression for $r_y$, when $M_1$ and $M_2$ lie in the same poloidal plane. In this case $\psi = 0$, and (\ref{expr:n_dot_s}) reduces to the form, which can be easily verified, Fig.~\ref{fig:2D_tokamak_sketch},
\begin{equation}
	\mathbf{n}\cdot\mathbf{s} = \frac{\left(R_2-R_1\right)\sin\theta+\left(Z_2-Z_1\right)\cos\theta}{\left[\left(R_2-R_1\right)^2 + \left(Z_2-Z_1\right)^2\right]^{1/2}}\equiv\sin\left(\theta+\chi\right), \label{expr:n_dot_s_reduced}
\end{equation}
where $\chi = \arctan[(Z_2-Z_1)/(R_2-R_1)]$.

\section*{Appendix B. Solution to the 3D heat equation for the elliptic laser beam}	
\renewcommand{\theequation}{B.\arabic{equation}}
\setcounter{equation}{0}
To obtain the general solution (\ref{expr:T_general}) and its approximation (\ref{expr:T_polynomial}), we briefly recall equations of the heat transport model. The thermal transfer is governed by the following linear equation,
\begin{eqnarray}
	\ddtp{T} = a^2 \left(\frac{\partial^2}{\partial{x}^2} + \frac{\partial^2}{\partial{y}^2} + \frac{\partial^2}{\partial{z}^2}\right)T + P, \label{eqn:T_app}
\end{eqnarray}
where $a^2 = \kappa/(\rho C_p)$,
\begin{equation}
	P(x,y,z,t) = P_0 \exp\left(-\frac{x^2}{r_x^2}-\frac{y^2}{r_y^2}\right) \delta(z) Y(t),
\end{equation}
$P_0 = 2I_0/(\rho C_p)$, $Y(t) = H(t_p - t)(1-\Delta{t/t_p})$, and $H(t)$ and $\delta(z)$ are the Heaviside and Dirac functions, respectively. The volumetric heat source is introduced to account for the sample heating by the laser irradiation and to subsequently remove the heat source from the boundary conditions, which are as follows
\begin{eqnarray}
	\frac{\partial{T}}{\partial{z}}(x,y,0,t) = 0, \qquad \lim\limits_{x,y,z\rightarrow\infty}T(x,y,z,t) = T_0, \label{eqn:BC_app}
\end{eqnarray}
Finally, the initial condition for the sample temperature is
\begin{equation}
	T(x,y,z,0) = T_0. \label{eqn:IC_app}
\end{equation}
The coordinates $xyz$ constitute the local Cartesian reference frame, as discussed in Appendix A.

By introducing the auxiliary variable, $u(x,y,z,t) = T(x,y,z,t) - T_0$, the problem is re-formulated as
\begin{eqnarray}
	\ddtp{u} = a^2 \left(\frac{\partial^2}{\partial{x}^2} + \frac{\partial^2}{\partial{y}^2} + \frac{\partial^2}{\partial{z}^2}\right)u + P, \label{eqn:u_app}\\
	\frac{\partial{u}}{\partial{z}}(x,y,0,t) = 0, \qquad \lim\limits_{x,y,z\rightarrow\infty}u(x,y,z,t) = 0,\\
	u(x,y,z,0) = 0. \label{eqn:u_IC}
\end{eqnarray}

The new system is solved by using the Green's function formalism \cite{sveshnikov2004lectures}. For the problem at hand, the three-dimensional Green function is given by
\begin{eqnarray}\label{expr:Green_function}
	G(x,y,z,\xi,\eta,\zeta,t-\tau) = \frac{1}{[2a\sqrt{\pi(t-\tau)}]^{3}}\nonumber\\
	\qquad\times
	\cases{
		\exp\left[-\frac{(x-\xi)^2 + (y-\eta)^2 + (z - \zeta)^2}{4a^2(t-\tau)}\right] \\
		+ \exp\left[-\frac{(x-\xi)^2 + (y-\eta)^2 + (z + \zeta)^2}{4a^2(t-\tau)}\right], \qquad t - \tau > 0, \\
		0, \qquad t - \tau \leq 0,
	}
\end{eqnarray}
and the general solution to (\ref{eqn:u_app})-(\ref{eqn:u_IC}) with respect to the sample temperature can be expressed as
\begin{eqnarray}
	T(x,y,z,t) = &T_0 + P_0 \int_{0}^{t} \dif{\tau}\frac{Y(\tau)}{\left[2a\sqrt{\pi(t-\tau)}\right]^3} 
	\nonumber\\
	&\int_{-\infty}^{+\infty}\dif{\xi}\dif{\eta} \exp\left[-\frac{(x-\xi)^2 + (y-\eta)^2 + z^2}{4a^2(t-\tau)}\right]\exp\left(-\frac{\xi^2}{r_x^2}-\frac{\eta^2}{r_y^2}\right). \label{expr:T_1}
\end{eqnarray}

To further simplify the relation, we expand the exponential factor $\exp(-\eta^2/r_y^2)$ as a convolutional integral,
\begin{equation}
	\exp\left(-\frac{\eta^2}{r_y^2}\right) = \int_{-\infty}^{+\infty}\dif\mu f(\mu) \exp\left[-\frac{(\eta - \mu)^2}{r_x^2}\right], \label{expr:transform}
\end{equation}
where
\begin{equation}
	f(\mu) = \frac{r_y}{r_x\sqrt{\pi\left(r_y^2 - r_x^2\right)}} \exp\left(-\frac{\mu^2}{r_y^2 - r_x^2}\right). \label{expr:f_conv}
\end{equation}
Note that the convolution (\ref{expr:transform}) is a must, if we are to obtain the analytically tractable relations expressed in terms of elementary functions, similar to the relation (20) of Ref.~\cite{stepanenko2024dimensional}. A straightforward step of transforming each exponent under the integral sign in (\ref{expr:T_1}), as is done in \cite{stepanenko2024dimensional}, is impractical, as it would lead to the functional series containing the incomplete elliptic integrals.

By introducing the dimensionless time, $\theta$, and the auxiliary radial coordinate, $\rho$,
\begin{equation}
	\theta = \frac{2a(t-\tau)^{1/2}}{r_x}, \qquad \rho^2 = x^2 + (y - \mu)^2,
\end{equation}
neglecting the dependence of (\ref{expr:T_1}) on the depth $z$ (due to arguments discussed in Sec.~\ref{sec:analytical}), we re-cast (\ref{expr:T_1}) as
\begin{equation}
	\frac{T(x,y,t)}{T_0} = 1 + J_0\int_{-\infty}^{+\infty}\dif\mu f(\mu)\int_{\theta_{\min}}^{\theta_{\max}} \dif\theta \frac{h(\theta)}{1+\theta^2}\exp\left(-\frac{\rho^2}{1+\theta^2}\right), \label{expr:T_2}
\end{equation}
where $J_0 = I_0 r_x / (T_0\kappa\sqrt{\pi})$, $h(\theta) = 1 - \Delta(\theta_{\max}^2 - \theta^2)/\theta_p^2$, $\theta_{\max} = 2a\sqrt{t}/r_x$, and $\theta_p = 2a\sqrt{t_p}/r_x$. The parameter $\theta_{\min} = 0$, if $t \leq t_p$, otherwise, $\theta_{\min} = 2a\sqrt{t-t_p}/r_x$. By substituting (\ref{expr:f_conv}) into (\ref{expr:T_2}), after some algebra one finds
\begin{equation}
	\frac{T(x,y,t)}{T_0} = 1 + J_0 \exp\left(-\frac{x^2}{r_x^2}-\frac{y^2}{r_y^2}\right) \mathcal{R}(x,y,t),
\end{equation}
where
\begin{equation}
	\mathcal{R}(x,y,t) = \int_{-\infty}^{+\infty}\dif\lambda\exp(-\lambda^2)\int_{\theta_{\min}}^{\theta_{\max}}\dif\theta \frac{h(\theta)}{1+\theta^2}\exp\left(\frac{\theta^2}{1+\theta^2}\rho^2\right),
\end{equation}
where we recall that $\rho^2 = x^2 + (y-\mu)^2 \equiv x^2 + \lambda^2$. By expanding the exponential factor under the integral sign into the Taylor series and performing integrations, we find, similarly to \cite{stepanenko2024dimensional},
\begin{equation}
	\mathcal{R}(x,y,t) = F(x,y,\theta_{\max}) - F(x,y,\theta_{\min}),
\end{equation}
where the function $F$ generalizes the analogous expression found in \cite{stepanenko2024dimensional} to the case of the  Gaussian beam with the elliptic shape,
\begin{equation}
	F(x,y,\theta) = \sum_{n=0}^{\infty}R_n(x,y)\left[K_1A_n(\theta) + K_2B_n(\theta)\right].
\end{equation}
In this expression, $K_1 = 1 - \Delta\left(\theta_{\max}/\theta_p\right)^2$, $K_2 = \Delta/\theta_p^2$,
\begin{eqnarray}
	A_n(\theta) = \frac{1}{n!} \int_{0}^{\theta} d\varphi \frac{\varphi^{2n}}{\left(1 + \varphi^2\right)^{n+1}},\\
	B_n(\theta) = \frac{1}{n!} \int_{0}^{\theta} d\varphi \frac{\varphi^{2n+2}}{\left(1 + \varphi^2\right)^{n+1}},
\end{eqnarray}
and the spatial factors $R_n(x,y)$ are given by
\begin{eqnarray}
	R_n(x,y) &= \frac{1}{\sqrt{\pi}} \int_{-\infty}^{+\infty} \dif\lambda \exp(-\lambda^2)\left[\xi^2 + \left(e\lambda - k\eta\right)^2\right]^n \\\nonumber
	&\equiv \sum_{m=0}^{n}\sum_{l=0}^{n-m} C_n^m C_{2n - 2m}^{2l} \frac{(2l-1)!!}{2^l} \xi^{2m} (k\eta)^{2n-2m-2l} e^{2l},
\end{eqnarray}
where $\xi = x/r_x$, $\eta = y/r_y$, $e=\sqrt{1-k^2}$ and $k = r_x/r_y$ are the eccentricity and inverse aspect ratio of the laser beam spot, correspondingly. In the circular beam limit $r_x = r_y = r_0$, we have $k = 1, e=0$, and the functions $R_n(x,y)$ reduce to $R_n(x,y) = (\xi^2 + \eta^2)^n = (r/r_0)^{2n}$, as discussed in Sec.~\ref{sec:analytical}. Explicit expressions for the first few approximants $R_n(x,y)$ are given in Appendix C.

\section*{Appendix C. Polynomial coefficients $A_n(\theta)$, $B_n(\theta)$, $R_n(x,y)$}
\renewcommand{\theequation}{C.\arabic{equation}}
\setcounter{equation}{0}
The first six triplets of the polynomial coefficients $A_n$, $B_n$, $R_n$ ($n = 0\div5$) that were used to reconstruct the temperature profile (\ref{expr:T_polynomial}) in Sec.~\ref{sec:analytical} are as follows
\begin{eqnarray}
	A_0(\theta) &= \arctan\theta,\\
	A_1(\theta) &= \frac12\arctan\theta - \frac{\theta}{2(\theta^2+1)},\\
	A_2(\theta) &= \frac{3}{16}\arctan\theta - \frac{\theta(5\theta^2 + 3)}{16(\theta^2+1)^2},\\
	A_3(\theta) &= \frac{5}{96}\arctan\theta - \frac{\theta(33\theta^4 + 40\theta^2 + 15)}{288(\theta^2+1)^3},\\
	A_4(\theta) &= \frac{35}{3072}\arctan\theta - \frac{\theta(279\theta^6+511\theta^4+385\theta^2+105)}{9216(\theta+1)^4},\\
	A_5(\theta) &= \frac{63}{30720}\arctan\theta - \frac{\theta(965\theta^8+2370\theta^6+2688\theta^4+1470\theta^2+315)}{153600(\theta+1)^5},\\
	B_0(\theta) &= \theta - \arctan\theta,\\
	B_1(\theta) &= \theta + \frac{\theta}{2(\theta^2+1)} - \frac32\arctan\theta,\\
	B_2(\theta) &= \frac{\theta}2 + \frac{\theta(9\theta^2 + 7)}{16(\theta^2+1)^2} - \frac{15}{16}\arctan\theta,\\
	B_3(\theta) &= \frac{\theta}6 + \frac{\theta(87\theta^4 + 136\theta^2 + 57)}{288(\theta^2+1)^3} - \frac{35}{96}\arctan\theta,\\
	B_4(\theta) &= \frac{\theta}{24} + \frac{\theta(325\theta^6 + 765\theta^4 + 643\theta^2 + 187)}{3072(\theta^2+1)^4} - \frac{315}{3072}\arctan\theta,\\
	B_5(\theta) &= \frac{\theta}{120} + \frac{\theta(4215\theta^8 + 13270\theta^6 + 16768\theta^4 + 9770\theta^2 + 2185)}{153600(\theta^2+1)^5}\\\nonumber &- \frac{693}{30720}\arctan\theta,\\
	R_0(x,y) &= 1,\\
	R_1(x,y) &= \xi^2 + k^2 \eta^2 + \frac12 e^2,\\
	R_2(x,y) &= \left(\xi^2 + k^2 \eta^2\right)^2 + e^2 \left(\xi^2 + 3k^2 \eta^2\right) + \frac34 e^4,\\
	R_3(x,y) &= 
	\left(\xi^2 + k^2\eta^2\right)^3 + 
	e^2 \left(\frac32 \xi^4 + 9 \xi^2 k^2 \eta^2 + \frac{15}{2}k^4\eta^4\right)\\\nonumber &+ 
	e^4 \left(\frac94 \xi^2 + \frac{45}{4}k^2\eta^2\right) + 
	\frac{15}{8}e^6,\\
	R_4(x,y) &= 
	\left(\xi^2 + k^2\eta^2\right)^4 + 
	e^2 \left(2\xi^6 + 18\xi^4 k^2\eta^2 + 30\xi^2 k^4\eta^4 + 14k^6\eta^6\right)\\\nonumber &+ 
	e^4 \left(\frac{9}{2}\xi^4 + 45\xi^2 k^2\eta^2 + \frac{105}{2}k^4\eta^4\right) + 
	e^6 \left(\frac{15}{2}\xi^2 + \frac{105}{2}k^2\eta^2\right) + 
	\frac{105}{16}e^8,\\
	R_5(x,y) &= \left(\xi^2 + k^2\eta^2\right)^5\\\nonumber & + 
	e^2 \left(\frac52 \xi^8 + 30\xi^6 k^2\eta^2 + 75\xi^4 k^4\eta^4 + 70\xi^2 k^6\eta^6 + \frac{45}{2}k^8\eta^8\right)\\\nonumber &+ 
	e^4 \left(\frac{15}{2}\xi^6 + \frac{225}{2}\xi^4 k^2\eta^2 + \frac{525}{2}\xi^2 k^4\eta^4 + \frac{315}{2}k^6\eta^6\right)\\\nonumber &+ 
	e^6 \left(\frac{75}{4}\xi^4 + \frac{525}{2}\xi^2 k^2\eta^2 + \frac{1575}{4}k^4\eta^4\right)\\\nonumber & + 
	e^8 \left(\frac{525}{16}\xi^2 + \frac{4725}{16}k^2\eta^2\right) +
	\frac{945}{32} e^{10}.
\end{eqnarray}

\bibliographystyle{ieeetr}
\bibliography{LID_elliptic_beam}
\end{document}